\documentstyle[12pt]{article}
\renewcommand
\baselinestretch{1.4}

\begin{document}

\def\beq{\begin{equation}}
\def\eeq{\end{equation}}
\def\bce{\begin{center}}
\def\ece{\end{center}}
\def\bea{\begin{eqnarray}}
\def\eea{\end{eqnarray}}
\def\ben{\begin{enumerate}}
\def\een{\end{enumerate}}
\def\ul{\underline}
\def\ni{\noindent}
\def\nn{\nonumber}
\def\bs{\bigskip}
\def\ms{\medskip}
\def\wt{\widetilde}
\def\brr{\begin{array}}
\def\err{\end{array}}
\def\dsp{\displaystyle}

\hfill UB-ECM-PF-94/7

\hfill March 8th, 1994

\vspace*{10mm}

\begin{center}

 {\large \bf AN EXTENSION OF THE CHOWLA-SELBERG FORMULA  \\
USEFUL
IN QUANTIZING WITH THE WHEELER-DE WITT EQUATION}

\vspace{4mm}

\renewcommand\baselinestretch{0.8}
\medskip

{\sc E. Elizalde}\footnote{E-mail: eli@ebubecm1.bitnet,
eli@zeta.ecm.ub.es} \\
Center for Advanced Studies (CEAB), C.S.I.C., 17300 Blanes \\
 and \ Department E.C.M. and I.F.A.E., Faculty of Physics,
University of  Barcelona, \\ Diagonal 647, 08028 Barcelona,
Catalonia, Spain

\vspace{25mm}

{\bf Abstract}

\end{center}

 The two-dimensional inhomogeneous
zeta-function series (with homogeneous part of the most
general Epstein type):
\[ \sum_{m,n \in \mbox{\bf Z}} (am^2+bmn+cn^2+q)^{-s},
\]  is analytically continued in the variable $s$ by using
zeta-function techniques. A simple formula is obtained, which
extends the Chowla-Selberg formula to inhomogeneous
Epstein zeta-functions. The new expression is then applied to
solve the problem of computing the determinant of the basic
differential operator that appears in an attempt at
quantizing gravity by using the Wheeler-De Witt equation in 2+1
dimensional spacetime with the torus topology.
\newpage

\section{Introduction}

In a recent publication \cite{carl}, dealing with the approach to
(2+1)-dimensional quantum gravity which consists in making direct
use of the Wheeler-De
Witt equation, Carlip has come across a rather involved
mathematical problem. Of course, none of the aproaches that have
been employed for the
quantization of gravity is simple, for different reasons. Here we
will concentrate only in the specific point of the whole problem
that has been risen by Carlip, and which concerns the calculation
of the basic determinant that appears in his method for the case
of the torus topology. It is the determinant corresponding to a
differential operator, $D_0$, which has the following
set of eigenfunctions and eigenvalues (for explicit details,
see \cite{carl} and references therein):
\beq
|mn> = e^{2\pi i (mx+ny)}, \hspace{20mm} l_{mn} =
\frac{4\pi^2}{\tau_2} |n - m \tau|^2 + V_0,
\label{espec}
\eeq
where $m$ and $n$ are integers, and $\tau$ and $\tau_2$
are the usual labels corresponding to the standard
two-dimensional metric for the torus
\beq
d\bar{s}^2 = \tau_2^{-1} |dx + \tau dy|^2,
\label{metr}
\eeq
with $x$ and $y$ angular coordinates of period 1 and $\tau =
\tau_1
+ i \tau_2$ the modulus (a complex parameter \cite{carl}). $V_0$
is the spatial integral of the relevant potential function
\cite{carl}.

At that point, the physical difficulty has boiled down to a well
formulated mathematical problem which, unfortunately, has no
straightforward solution from, e.g. the zeta functions which
commonly
appear in  physical or mathematical references. Since let us
recall that, in fact, the best way to
obtain the determinant once the spectrum of the operator is known
is through the calculation of the corresponding zeta function,
$\zeta_{D_0}$. After simplifying the notation a little, one
easily recognizes that one has to deal here with a series of the
form
\begin{equation}
F(s;a,b,c;q) \equiv  {\sum_{m,n \in \mbox{\bf Z}}}'
(am^2+bmn+cn^2+q)^{-s},
\label{1}
\end{equation}
the prime meaning that the term with both $m=n=0$ is absent from
the sum. Of
course this distinction needs not to be done when $q\neq 0$ (the
value of such term being then trivially $q^{-s}$), but it is
certainly
important
for considering the particular case $q=0$ (see later). One is
interested in obtaining the function $F(1;a,b,c;q)$
of $a,b,c,q$, since this expression comes inside a functional
integral
which involves the relevant variables of the problem. As it
stands, Eq. (\ref{1}) has no sense
for $s=1$, and it is also clear that analytic continuation to
such value of $s$ hits a pole and, therefore, must be
conveniently defined. This has been done successfully in the
literature (see \cite{1}).

In what follows we will calculate the sum (\ref{1}) ---and its
corresponding analytical continuat\-ion--- and also its
derivative
with respect to $s$, by means of zeta-function techniques.
The final expresion will be remarkably simple,  involving just
(apart from finite sums) a quickly convergent series of
exponentially vanishing integrals (of Bessel function type). It
is, in fact, a generalization of the
celebrated (by the mathematicians) Chowla-Selberg formula. We
will
proceed step by step, starting from some particular, more simple
cases.

The case $b=0$ will be treated in Sect. 2, and the general
homogeneous case ($q=0$) in Sect. 3 ---this is the Chowla-Selberg
formula itself. In Sect. 4 we will derive the new formula, which
is capable of dealing with the general situation ($q\neq 0$). The
explicit use of the formula in the quantization of
(2+1)-dimensional gravity will be treated in Sect. 5, where
specific results for this physical application will be given. For
the sake of comparison, we present in Sect. 6 an alternative
treatement by means of Eisenstein series. Finally, Sect. 7 is
devoted to conclusions.
 \medskip

\section{Case b=0}

Remember that
\begin{equation} \zeta(s) = \sum_{n=1}^\infty \frac{1}{n^s}, \ \
\  \
\zeta_H(s,a) = \sum_{n=0}^\infty \frac{1}{(n+a)^s}, \ \ \ \
a\neq 0,-1,-2,\ldots, \ \ \ \ Re \ s > 1,
\end{equation}
define the Riemann and Hurwitz zeta functions, respectively,
which can be analytically continued in $s$ as  meromorphic
functions to the whole complex $s$-plane (with just a simple pole
at $s=1$). In general, in zeta-function regularization
 we are done when the
result can be expressed in terms of these simple functions.
However, in
the more difficult case involved here, we will have a different
function
as the most elementary one (see Eq. (\ref{zeh1}) below).

Considering again the general series (\ref{1}), the parenthesis
in this expression must be visualized as an
inhomogeneous quadratic form:
\begin{equation}
Q(x,y)+q, \ \ \ \  Q(x,y)  \equiv ax^2+bxy+cy^2,
\end{equation}
restricted to the  integers.
We shall assume all the time that $b>0$, that the discriminant
\beq
\Delta =4ac-b^2 >0,
\eeq
and that $q$ is such that $Q(m,n)+q \neq 0$, $\forall \, m,n \in
$ {\bf Z}. In terms of the corresponding
physical constants, as they appear in \cite{carl}, these
conditions are
indeed satisfied (for the physically relevant cases).
We start by studying some particular situations which, for the
benefit of the reader, may be interesting to recall.

The case $b=0$ corresponds to a situation that we have considered
in former papers (even in more general terms) and for which we
have already derived explicit formulas
\cite{1} (see also \cite{2}). In particular, for the series
\begin{equation}
E_2^c(s;a_1,a_2;c_1,c_2) \equiv
\sum_{n_1, n_2=0}^{\infty} \left[ a_1(n_1+c_1)^2
+a_2(n_2+c_2)^2 +c\right]^{-s}.
\end{equation}
we have obtained the following expression:
\begin{eqnarray}
&& E_2^c (s;a_1,a_2;c_1,c_2) =
\frac{a_2^{-s}}{\Gamma (s)} \sum_{m=0}^{\infty} \frac{(-1)^m
\Gamma
(s+m)}{m!} \left( \frac{a_1}{a_2} \right)^m \zeta_H (-2m,c_1)
\nonumber  \\
&& \hspace{2cm} \times  E_1^{c/a_2} (s+m;1;c_2) +
 \frac{a_2^{1/2-s}}{2} \sqrt{\frac{\pi}{a_1}}\,
\frac{\Gamma
\left( s- \frac{1}{2} \right) }{\Gamma (s) } E_1^{c/a_2} (s-
1/2;1;c_2) \nonumber  \\
&& \hspace{1cm} +  \frac{2\pi^s}{\Gamma (s)}  a_1^{-s/2-1/4}
a_2^{-s/2+1/4} \sum_{n_1=1}^{\infty} \sum_{n_2=0}^{\infty}
n_1^{s- 1/2} \cos (2\pi n_1 c_1) \left[  a_2 (n_2+c_2)^2+c
\right]^{-s/2+1/4} \nonumber \\
&& \hspace{2cm} \times
K_{s-1/2} \left(
 \frac{2\pi n_1}{\sqrt{a_1}} \sqrt{  a_2 (n_2+c_2)^2+c }
\right), \label{5}
\end{eqnarray}
where
\begin{eqnarray}
 E_1^c (s;a_1;c_1) &=&
\frac{c^{-s}}{\Gamma (s)} \sum_{m=0}^{\infty} \frac{(-1)^m
\Gamma
(s+m)}{m!} \left( \frac{a_1}{c} \right)^m \zeta_H (-2m,c_1) +
 \frac{c^{1/2-s}}{2} \sqrt{\frac{\pi}{a_1}}\,
\frac{\Gamma
\left( s- \frac{1}{2} \right) }{\Gamma (s) } \nonumber  \\
&+& \frac{2\pi^s}{\Gamma (s)}\, a_1^{-s/2-1/4}
c^{-s/2+1/4} \sum_{n_1=1}^{\infty}n_1^{s-1/2} \cos (2\pi n_1c_1)
K_{s-1/2} \left(
2\pi n_1  \sqrt{\frac{c}{a_1}} \right), \label{6}
\end{eqnarray}
and being $K_\nu$ the modified Bessel functions of the second
kind.
In order to specify the formula por the present case, we just
have to substitute: $c_1=c_2=1$ (the Hurwitz zeta functions
 turn simply into ordinary Riemann ones), $a_1=a$, $a_2 =c$ and
$c=q$, and be a bit careful with the summation
range; but this can be easily taken care of (see, for instance,
\cite{3} where the explicit formulas relating the cases of doubly
infinite summation ranges and simply infinite ranges are given).
\medskip

\section{Case q=0}
Here, the Chowla-Selberg formula \cite{4} for the (general
homogeneous) Epstein
zeta function \cite{eps} corresponding to the quadratic form $Q$
is to be used. (This is an expression well known
in number theory \cite{dic} but not so much in mathematical
physics.) The result is
\begin{eqnarray}
&& F(s;a,b,c;0) = 2\zeta (2s)\, a^{-s} + \frac{2^{2s}
\sqrt{\pi}\, a^{s-1}}{\Gamma (s) \Delta^{s-1/2}} \,\Gamma (s
-1/2) \zeta (2s-1) + \frac{2^{s+3/2} \pi^s }{\Gamma (s)
\Delta^{s/2-1/4}\sqrt{a}}
\nn \\ && \hspace{1cm} \times \sum_{n=0}^\infty n^{s-1/2}
 \sigma_{1-2s}(n) \cos (n\pi b/ a) \int_0^\infty dt \, t^{s-3/2}
\exp \left[ -\frac{\pi n \sqrt{\Delta}}{2a} (t+t^{-1}) \right],
\label{cs1}
\end{eqnarray}
where
\begin{equation}
\sigma_s(n) \equiv \sum_{d|n} d^s,
\end{equation}
namely the sum over the $s$-powers of the divisors of $n$.

This formula is very useful and its practical application quite
simple. In fact, the two first terms are just nice, while
the last one (impressive in appearence) is very quickly
convergent and thus absolutely harmless in practice: only a few
first terms of the series need to be
calculated, even if one needs excepcionally good accuracy. One
should also notice that the pole of $F$ at $s=1$ appears through
$\zeta (2s-1)$ in the second term, while for $s=1/2$, the
apparent singularities of the first and second terms cancel each
other and no pole is formed.

A closer, quantitative idea about the integral can be got from
the following closed expression for it:
\beq
\int_0^\infty dt \, t^{\nu -1}
\exp \left( -\frac{\alpha}{t}- \beta t \right) =
2 \left( \frac{\alpha}{\beta} \right)^{\nu /2} K_\nu \left( 2
\sqrt{\alpha \beta} \right),
 \label{inbe1}
\eeq
$K_\nu$ being again the modified Bessel function of the second
kind.
In particular, by calling the integral
\beq
I(n,s) \equiv \int_0^\infty dt \, t^{s-3/2}
\exp \left[ -\frac{\pi n \sqrt{\Delta}}{2a} (t+t^{-1}) \right],
\label{intg1}
\eeq
one has
\bea
&& I(n,0) = \sqrt{\frac{2a}{n\sqrt{\Delta}}} \, \exp \left(
-\frac{\pi n \sqrt{\Delta}}{a} \right) =I(n,1), \ \ \ \ I(n,1/2)
=
2 K_0(\pi n \sqrt{\Delta}/a), \nn  \\ && I(n,2) = \frac{a+ \pi n
\sqrt{\Delta}}{\pi n\sqrt{\Delta}}
\sqrt{\frac{2a}{n\sqrt{\Delta}}}
\, \exp \left( -\frac{\pi n \sqrt{\Delta}}{a} \right), \nn \\ &&
I(n,3) = \frac{3a^2+3 \pi n a \sqrt{\Delta}+ \pi^2 n^2
\Delta}{\pi^2 n^2 \Delta} \sqrt{\frac{2a}{n\sqrt{\Delta}}} \,
\exp
\left( -\frac{\pi n \sqrt{\Delta}}{a} \right)
\label{intg2}
\eea
As functions of $n$, all these expressions
share the common feature of being exponentially decreasing with
$n$.

\medskip

\section{The general case a,b,c,q $\neq$ 0}

This case is more difficult. To handle it, we can choose to go
through the
whole derivation of the Chowla-Selberg formula for the
quadratic form $Q$ and see the differences
introduced by the inhomogeneity (the constant $q$). Instead, we
will here undertake a more down-to-earth derivation, which will
be similar to the
ones that we have successfully employed several times in former
papers
---in particular to obtain Eqs. (\ref{5}) and (\ref{6}). Since
the
technicalities of the method have been abundantly discussed
before
\cite{1} (see also \cite{3}), we will here consider the main
steps of the proof only. They are the following.
(i) Rewrite the initial
expression (\ref{1}), (\ref{5}), by using the gamma function
identity
\beq
\sum_{m,n} (Q+q)^{-s} = \frac{1}{\Gamma (s)} \sum_{m,n}
\int_0^\infty du \, u^{s-1} e^{-(Q+q)u}.
\eeq
(ii) Expand the exponential in terms of power series of $m$ and
$n$ and interchange the order of the summations, i.e., the sum
over such expansion  with the sums over $m$
and $n$ or ---equivalently in this case--- use Jacobi's theta
function fundamental identity (as
can be found, for instance, in \cite{erd}). The equivalence of
both methods was explicitly proven in \cite{elif2}. The second
one starts here from a trivial rewriting of the non-negative
quadratic form $Q(m,n)$ as the sum of two squares
\beq
Q(m,n)= a\left[ \left( m + \frac{bn}{2a} \right)^2 +
\frac{\Delta}{4a^2} n^2 \right],
\label{sts}
\eeq
and proceeds by considering the summation over $m$,
while treating first $n$ as a parameter.
 (iii) Finally, make the following change of variables (for
convenience)
\beq
u=\frac{2\pi m}{\sqrt{\Delta}\, n} t
\eeq
and use the same idea as in Eq.  (\ref{cs1}) of rewriting the
double sum as a sum over the product $mn$ and (a finite one) over
the divisors of the product:
\beq
\sum_{n_1,n_2} \left( \frac{n_1}{n_2} \right)^{s-1/2} =
\sum_{n_1,n_2} \left( n_1n_2 \right)^{s-1/2} n_2^{1-2s} =
\sum_n n^{s-1/2}  \sum_{d|n} d^{1-2s}
\eeq
(this factor appears when the change of variables is performed).
On the other hand, the term $\xi_n \equiv bn/(2a)$ in the first
square
of the decomposition (\ref{sts}) (that may be written
$(m+\xi_n)^2
+n^2$) leads to a cosine factor in the final expression, e.g.
\beq
 \cos (2\pi \xi_n) = \cos (n\pi b/ a).
\eeq
This is also explained in detail in \cite{1} (but notice the
small
mistake in the first of these references, that was later
corrected in the subsequent ones).

 By doing all this,
the following generalized expression is obtained
\begin{eqnarray}
F(s;a,b,c;q) &=&  {\sum_{m,n \in \mbox{\bf Z}}}' [Q(m,n)+q]^{-s}
= {\sum_{m,n \in \mbox{\bf Z}}}' (am^2+bmn+cn^2+q)^{-s} \nn \\ &=
&  2\zeta_{EH} (s,4aq/\Delta)\, a^{-s} + \frac{2^{2s}
\sqrt{\pi}\, a^{s-1}}{\Gamma (s) \Delta^{s-1/2}} \, \Gamma (s -
1/2) \zeta_{EH} (s-1/2,4aq/\Delta) \nn \\ && + \frac{2^{s+3/2}
\pi^s }{\Gamma (s)
\Delta^{s/2-1/4}\sqrt{a}}
\sum_{n=0}^\infty
n^{s-1/2} \cos (n \pi b/a)  \sum_{d|n} d^{1-2s}
\int_0^\infty dt \, t^{s-3/2} \nn \\ && \hspace{1cm} \times \exp \left\{
-\frac{\pi n \sqrt{\Delta}}{2a}
\left[ \left( 1+ \frac{4aq}{\Delta d^2} \right) t +
t^{-1} \right] \right\},
\label{gcso}
\end{eqnarray}
where the function $\zeta_{EH} (s,p) $ (one dimensional
Epstein-Hurwitz or inhomogeneous Epstein) is given by
 \begin{eqnarray}
 \zeta_{EH}(s;p) &=& \sum_{n=1}^\infty  \left( n^2 + p
\right)^{-s} \label{zeh1} \\
& =& -\frac{p^{-s}}{2} + \frac{\sqrt{\pi} \, \Gamma (s-
1/2)}{2\, \Gamma (s)}
p^{-s+1/2} + \frac{2\pi^s p^{-s/2 +1/4}}{\Gamma (s)}
\sum_{n=1}^\infty n^{s -1/2} K_{s -1/2} (2\pi n\sqrt{p}),
\nonumber
\end{eqnarray}
and is studied in full detail in \cite{elif3} (with numerical
tables, plots, and a couple of explicit physical applications).

It is remarkable that the integral inside the series of the
new expression can still be written in a closed form using
(\ref{inbe1})
---as in the case of Eq. (\ref{cs1}).
Calling now the integral
\beq
J(n,s) \equiv \int_0^\infty dt \, t^{s-3/2}
\exp \left\{ -\frac{\pi n \sqrt{\Delta}}{2a} \left[ \left( 1+
\frac{4aq}{\Delta d^2} \right)t+t^{-1}\right] \right\},
\label{intg3}
\eeq
we obtain, in particular,
\bea
&& J(n,0) = \sqrt{\frac{2a}{n\sqrt{\Delta}}} \, \exp \left[
-\frac{\pi n}{a} \left( \Delta + \frac{4aq}{d^2}
\right)^{1/2} \right], \nn \\ && J(n,1/2) =
2 K_0\left( \frac{\pi n }{a}
\sqrt{ \Delta + \frac{4aq}{d^2}}
\right), \nn \\ &&  J(n,1) =
\sqrt{\frac{2a \sqrt{\Delta}}{n}} \left( \Delta + \frac{4aq}{d^2}
\right)^{-1/2} \, \exp \left[
-\frac{\pi n}{a} \left( \Delta + \frac{4aq}{d^2}
\right)^{1/2} \right], \nn \\ && J(n,2) = \left(\frac{a}{\pi n} +
\sqrt{\Delta + \frac{4aq}{d^2}} \right)
\sqrt{\frac{2a \Delta^{3/2}}{n}} \left( \Delta + \frac{4aq}{d^2}
\right)^{-3/2} \nn \\ && \hspace{2cm} \times \exp \left[
-\frac{\pi n}{a} \left( \Delta
+ \frac{4aq}{d^2} \right)^{1/2} \right], \nn \\ &&
J(n,3) = \left( \frac{3a^2}{\pi^2 n^2} + \frac{3 a}{\pi n}
\sqrt{\Delta + \frac{4aq}{d^2}}+
\Delta + \frac{4aq}{d^2} \right) \,
\sqrt{\frac{2a \Delta^{5/2}}{n}}
\nn \\ && \hspace{2cm} \times
 \left(  \Delta +
\frac{4aq}{d^2} \right)^{-5/2}\, \exp
\left[ -\frac{\pi n}{a} \left(  \Delta + \frac{4aq}{
d^2} \right)^{1/2}  \right],
\label{intg4}
\eea
which are again exponentially decreasing with $n$.

Expression (\ref{gcs}) itself can be written also in terms of
these Bessel functions:
 \beq
 F(s;a,b,c;q) =
  2\zeta_{EH} (s,4aq/\Delta)\, a^{-s} + \frac{2^{2s}
\sqrt{\pi}\, a^{s-1}}{\Gamma (s) \Delta^{s-1/2}} \, \Gamma (s -
1/2) \zeta_{EH} (s-1/2,4aq/\Delta)
\label{gcs} \eeq
\[ +
\frac{8 \, (2\pi)^s }{\Gamma (s) \Delta^{s-1/2}\sqrt{2a}}
\sum_{n=0}^\infty
n^{s-1/2} \cos (n \pi b/a) \sum_{d|n} d^{1-2s}
 \left( \Delta + \frac{4aq}{d^2} \right)^{s/2-1/4}
K_{s - 1/2}\left( \frac{\pi n}{a}
\sqrt{ \Delta + \frac{4aq}{d^2}} \right). \]
Eq. (\ref{gcs}) is the fundamental result of this paper and must
be given a name. We propose to call it {\it
inhomogeneous} or {\it generalized Chowla-Selberg} formula. To
our
knowledge, it has never appeared before in the mathematical (or
physical) literature.

\medskip

\section{Explicit application of the formula in
quantizing gravity through the Wheeler-De Witt equation}

As discussed in \cite{carl}, the quantization of
gravity in 2+1 dimensions by means of the Wheeler-De Witt
equation, in a spacetime with the topology  {\bf R}$\times${\bf
T}$^2$ ({\bf T}$^2$ being the two-dimensional
torus), of standard metric
given by (\ref{metr}), proceeds through the calculation of the
zeta function corresponding to the basic differential operator
$D_0$, which has a spectral decomposition given by (\ref{espec}).
In terms of the function $F(s;a,b,c;q)$ (\ref{1}), the zeta
function of $D_0$ is
\beq
\zeta_{D_0} (s) =F\left(s; 4\pi^2/\tau_2, -8\pi^2 \tau_1 /\tau_2,
4\pi^2 (\tau_1^2+\tau_2^2)/\tau_2; V_0\right).
\eeq
One has, in particular, $\Delta =64\pi^4$ and using Eq.
(\ref{gcs}) one gets
\bea
\zeta_{D_0} (s)& =& \frac{2^{-2s+1}\pi^{-2s}}{\tau_2^{-s}}
\zeta_{EH} \left(s,V_0/(4\pi^2\tau_2)\right) \nn \\ && +
\frac{2^{-
2s+1}\pi^{-2s+1/2}\Gamma (s-1/2)}{\tau_2^{s-1} \Gamma (s)}
\zeta_{EH} \left(s-1/2,V_0/(4\pi^2\tau_2)\right) \nn \\ && +
 \frac{2^{-2s+2}\pi^{-s}\sqrt{\tau_2}}{ \Gamma (s)}
\sum_{n=0}^\infty
n^{s-1/2} \cos (2n\pi \tau_1)  \sum_{d|n} d^{1-2s}
\int_0^\infty
dt \, t^{s-3/2} \nn \\ &&  \hspace{2cm} \times \exp \left\{ -n\pi
\tau_2
\left[ \left( 1+ \frac{V_0}{4\pi^2 d^2\tau_2} \right) t +
t^{-1} \right] \right\},
\label{gcs1}
\end{eqnarray}
with
 \bea
&& \zeta_{EH}\left(s;V_0/(4\pi^2\tau_2)\right) =  -2^{2s-
1}\pi^{2s}\left(\frac{V_0}{\tau_2} \right)^{-s} +2^{2s-2}\pi^{2s-
1/2} \frac{\Gamma (s- 1/2)}{\Gamma (s)} \left(\frac{V_0}{\tau_2}
\right)^{-s+1/2}\nn \\ && \hspace{15mm} + \frac{2^{s+1/2}\pi^{2s-
1/2}}{\Gamma (s)} \left(\frac{V_0}{\tau_2} \right)^{-s/2 +1/4}
\sum_{n=1}^\infty n^{s -1/2} K_{s -1/2} \left( n\sqrt{V_0/\tau_2}
\right).
\end{eqnarray}

The quantity of interest is the determinant of the operator $D_0$
(see \cite{carl}). This is most conveniently computed by means of
its
zeta function. In particular:
\beq
{\det}^{1/2}D_0= \exp \left[ - \frac{1}{2} \zeta_{D_0}'(0)
\right].
\eeq
Thus, we must now calculate the derivative of (\ref{gcs}) at
$s=0$. We have, for the general function $F(s;a,b,c;q)$
\bea
F'(0;a,b,c;q) &=& \ln a + 2\zeta_{EH}' (0,4aq/\Delta) -
\frac{2\pi
\sqrt{\Delta}}{a} \zeta_{EH} (-1/2,4aq/\Delta)  \nn \\ && + 4
\sum_{n=1}^\infty
n^{-1} \cos (n\pi b/a) \sum_{d|n} d \,
\exp \left[
-\frac{\pi n}{a} \left( \Delta + \frac{4aq}{d^2}
\right)^{1/2} \right],
\label{gcsp1}
\end{eqnarray}
where
\beq
\zeta_{EH}' (0;p) = - \pi \sqrt{p} + \frac{1}{2} \ln  p +
2p^{1/4}  \sum_{n=1}^\infty n^{-1/2}  K_{1/2} (2n\pi  \sqrt{p}),
\label{eh11}
\eeq
while for $ \zeta_{EH} (-1/2;p)$ the principal part prescription
(PP)
is to be used (see \cite{1,7,8}):
\beq
PP \zeta_{EH} (-1/2;p) = - \frac{p}{4} - \frac{\sqrt{p}}{2} -
\frac{\sqrt{p}}{\pi} \sum_{n=1}^\infty n^{-1}  K_1 (2n\pi
\sqrt{p}).
\label{eh22}
\eeq
Finally, for the determinant of $D_0$, we obtain
\bea
{\det}^{1/2} D_0 &=& \exp \left\{ - \frac{1}{2} \ln a -
\zeta_{EH}'
\left(0;V_0/(4\pi^2\tau_2)\right) + 2 \pi \tau_2 \zeta_{EH}
\left(-1/2;V_0/(4\pi^2\tau_2)\right)\right.\nn \\ && \left. - 2
\sum_{n=1}^\infty n^{-1} \cos (2n\pi \tau_1 )  \sum_{d|n} d \
 \exp \left[ -2n\pi  \tau_2 \left( 1+
\frac{V_0}{4\pi^2\tau_2d^2} \right)^{1/2} \right]\right\},
\label{gcsp2}
\end{eqnarray}
with $ \zeta_{EH}' \left(0;V_0/(4\pi^2\tau_2)\right)$ and $
\zeta_{EH} \left(-1/2;V_0/(4\pi^2\tau_2)\right)$ being given by
expressions (\ref{eh11}) and  (\ref{eh22}) above, putting
$p =V_0/(4\pi^2\tau_2)$. This yields
\bea
{\det}^{1/2} D_0 &=& \frac{\tau_2}{\sqrt{V_0}} \, \exp \left\{
- \frac{V_0}{8\pi} + \frac{1}{2} \sqrt{\frac{V_0}{\tau_2}} -
\frac{1}{2} \sqrt{\tau_2 V_0} \right. \nn \\ && -
\sqrt{\frac{2}{\pi}} \left( \frac{V_0}{\tau_2}\right)^{1/4}
\sum_{n=1}^\infty n^{-1/2}  K_{1/2} \left( n
\sqrt{\frac{V_0}{\tau_2}} \right)  - \frac{1}{\pi} \sqrt{\tau_2
V_0}  \sum_{n=1}^\infty n^{-1}  K_1 \left( n
\sqrt{\frac{V_0}{\tau_2}} \right)
\nn \\ && \left. - 2
\sum_{n=1}^\infty n^{-1} \cos (2n\pi \tau_1)  \sum_{d|n} d \
 \exp \left[ -2n\pi  \tau_2 \left( 1+
\frac{V_0}{4\pi^2\tau_2d^2} \right)^{1/2} \right]\right\},
\label{gcsp3}
\end{eqnarray}

We observe again that the final formula is really simple since,
in practice, it provides a very good approximations with just a
few terms, which are, on its turn, elementary functions of the
relevant variables and parameters. This is so, because the
infinite series that appear converge extremely quickly (terms
exponentially decreasing with $n$). In an asymptotical approach
to the determinant, only the first line in Eq. (\ref{gcsp3}) is
relevant (as we will show) and the three series can be
eliminated altogether.

{}From the detailed analysis in \cite{carl}, it follows that the
quantity to be calculated now is the derivative with respect to
$V_0$ of the above determinant, since this quantity vanishes
precisely at the solutions of the Hamiltonian constraint (always
in the language of quantization through the corresponding
Wheeler-De Witt equation). In other words, the solutions of the
equation
\beq
\frac{\partial}{\partial V_0} \, {\det}^{1/2} D_0  =0,
\label{ee1}
\eeq
will yield the conditions that the quantized magnitudes and
parameters are bound to satisfy as a consequence of the
Wheeler-De Witt equations. If this does not provide all the
solutions of such (very involved) differential equations, at
least gives us important clues about their behaviour (this is one
of the two basic problems of the approach in \cite{carl}, namely
that of understanding the determinant of $D_0$). After some
calculations one finds that Eq. (\ref{ee1}) can be written as
\bea
&& {\det}^{1/2} D_0 \times \left[ - \frac{1}{8 \pi} - \frac{1}{2V_0} -
\frac{1}{4} \, \sqrt{\frac{\tau_2}{V_0}} + \frac{1}{4
\sqrt{\tau_2 V_0}} \right. \nn \\ && -
\frac{(\tau_2/V_0)^{3/4}}{2 \sqrt{2\pi} \, \tau_2}
\sum_{n=1}^\infty n^{-1/2}  K_{1/2} \left( n
\sqrt{\frac{V_0}{\tau_2}} \right)  - \frac{1}{2\pi} \sqrt{
\frac{\tau_2}{V_0}} \sum_{n=1}^\infty n^{-1}  K_1 \left( n
\sqrt{\frac{V_0}{\tau_2}} \right)  \nn \\ && -
\frac{(\tau_2/V_0)^{1/4}}{\sqrt{2\pi} \, \tau_2}
\sum_{n=1}^\infty n^{1/2}  {K_{1/2}}' \left( n
\sqrt{\frac{V_0}{\tau_2}} \right)  - \frac{1}{2\pi}
\sum_{n=1}^\infty {K_1}' \left( n \sqrt{\frac{V_0}{\tau_2}}
\right)
\label{ee2}
 \\ && \left. + \frac{1}{2\pi}
\sum_{n=1}^\infty \cos (2n\pi \tau_1) \sum_{d|n}
 \left( d^2+ \frac{V_0}{4\pi^2\tau_2} \right)^{-1/2}
\exp \left( -2n\pi  \tau_2 \sqrt{ 1+
\frac{V_0}{4\pi^2\tau_2d^2}} \right)\right] =0 \nn
\end{eqnarray}
(the primes mean here derivatives of the Bessel functions).
In principle ---the consistency of the approximation is to be
checked {\it a posteriori}--- Eq.  (\ref{ee2}) can be reduced to
the very simple expression
\beq
\frac{1}{2\pi} \, \frac{V_0}{\tau_2} + \left( 1 - \frac{1}{\tau_2}
\right) \sqrt{\frac{V_0}{\tau_2}} + \frac{2}{\tau_2} \simeq 0.
\label{seq1}
\eeq
The analysis of this last equation is easy to do. From its discriminant
it turns out  that real solutions can only be obtained when
\beq
\tau_2 \leq \tau_2^{(1)} \ \ \ \ \mbox{or} \ \ \ \  \tau_2 \geq
\tau_2^{(2)}, \label{2sol}
\eeq
with
\beq
\tau_2^{(1)} \equiv 1 + \frac{2}{\pi} \left( 1 - \sqrt{\pi +1}
\right) = 0.34104103, \ \  \  \tau_2^{(2)} \equiv 1 + \frac{2}{\pi}
\left( 1 + \sqrt{\pi +1} \right) = 2.93219852.
\label{2solp}
\eeq
Moreover, the special situation when $\tau_2 =1$ leads to the
non-physical result $V_0=-4\pi$.
Now, from the constraint $\sqrt{V_0/\tau_2} >0$ (in order the
whole approximation to have sense) we see that the second possibility
in (\ref{2sol}) just disappears. For the first, when $\tau_2 \rightarrow
0$ it turns out that $\sqrt{V_0/\tau_2} \sim 2\pi/\tau_2   \rightarrow
\infty $ or  $\sqrt{V_0/\tau_2}  \rightarrow 2$. When $\tau_2$
moves within the allowed range $0 < \tau_2 \leq \tau_2^{(1)}$ (the first
interval of (\ref{2sol})), the corresponding two solutions
$\sqrt{V_0/\tau_2}$ of the quadratic equation (\ref{seq1})
sweep the following intervals
\beq
0 < \tau_2 \leq \tau_2^{(1)} = 0.34 \longrightarrow \left\{ \brr{c} +
\infty > \sqrt{V_0/\tau_2} \geq 2\pi \frac{\dsp\sqrt{\pi +1} -1}{\dsp\pi
+2 - 2 \sqrt{\pi +1}} = 6.07, \\  2 < \sqrt{V_0/\tau_2} \leq 2\pi
\frac{\dsp\sqrt{\pi +1} -1}{\dsp\pi +2 - 2 \sqrt{\pi +1}} = 6.07, \err
\right.
\label{2sol2}
\eeq
in the order indicated, respectively. It is easy to check that the first
of these
two alternatives (corresponding to the bigger root of (\ref{seq1}))
provides an absolutely consistent approximate solution to the exact
equation (\ref{ee2}). On the contrary, the second alternative
(corresponding to the smaller root of (\ref{seq1})) does not actually
provide a consistent approximate solution (unless $\tau_2$ is
close to $\tau_2^{(1)}$).
Direct investigation of other possible roots of Eq. (\ref{ee2}) (e.g.,
involving some terms of the series) is a quite difficult issue.
 \medskip

\section{An alternative treatement by means of Eisenstein
series}

An alternative way of treating the general case is the
following (see \cite{carl}).
 The inhomogeneity (the $q$ term here) is
taken care of by the simplest (but hardly economic) method of
performing
a binomial expansion of the sort \cite{8}
\begin{equation}
\sum_{k=0}^\infty \frac{\Gamma (s+k)}{k!\, \Gamma (s)} q^k E(z,
s+k),
\end{equation}
where $E(z,s)$ is an Eisenstein series (see, for instance, Lang
\cite{5} or Kubota \cite{6}), which is obtained from
$F(s;a,b,c;0)$ by
doing the substitution
\begin{equation}
2z = a+i u, \ \ \ \  c= C \frac{u}{2},
\end{equation}
so that
\begin{equation}
E(z,s) = {\sum_{m,n=0}^\infty}' \ (u/2)^s |m+n z|^{-2s},
\end{equation}
and has the series expansion
\begin{eqnarray}
E(z,s) &=& 2 \zeta (2s) + 2 \sqrt{\pi}\, (u/2)^{1-s}
\frac{\Gamma (1-s/2) \zeta (2s-1)}{\Gamma (s)} \nonumber \\
&& +2 \sum_{m=1}^\infty \sum_{n\neq 0} e^{i \pi mn a} \left(
\frac{2 |n|}{m u} \right)^{s-1/2} K_{s-1/2} (\pi mnu /2).
\end{eqnarray}

It is important to notice, however, that when doing things in
this last way the final result is expressed in terms of {\it
three} infinite sums, while in the first general procedure only
{\it
one} infinite sum appears (together with a finite sum, for every
index $n$, over the divisors of $n$), and it is very quickly
convergent.
Notice, moreover,
how the $d$-term in the exponent in Eq. (\ref{gcs}), when
expanded
in power series, gives rise to the binomial sum corresponding to
the
last treatement. The advantage of the use of the
 method developed in
Sect. 5, steming from Eq. (\ref{gcs}), seems clear (expanding a
negative exponential is in general computationally disastrous).
\medskip

\section{Conclusions}

The main results of this paper have been the derivation of
equation
(\ref{gcs}) which generalizes the Chowla-Selberg formula
(\ref{cs1})
and its physical application to calculate the determinant
(\ref{gcsp2}) and its derivative, Eqs. (\ref{ee2}) and
(\ref{seq1}). To
have at our disposal an exact expression for dealing with
inhomogeneous
Epstein zeta functions is certainly an interesting thing from the
physical point of view, since this kind of zeta functions appear
frequently in modern applications that are no more restricted to
zero
temperature or Euclidean spacetime. The simplicity of the result
is
remarkable. Namely, the fact that, in practice, we just need
a couple of
terms of the formula (\ref{gcs}) even if we want to obtain very
accurate numerical results.

Two problems were singularized out in Ref. \cite{carl}
as the main difficulties that appear in the quantization of
(2+1)-dimensional gravity through the Wheeler-De Witt equation:
(a) to give grounds for the choice of the specific operator
ordering of the Hamiltonian constraint which leads to the Wheeler-De
Witt equation of the quantized system, and (b) to understand
the functional dependence of the determinant  $\det^{1/2} D_0$ in terms
of the relevant variables and to obtain its extrema as a function of
the potential $V_0$.
In the present paper we have been able to solve problem (b)
---at least partially--- by means of a numerically consistent
approximation. Furthermore, we have now also the possibility,
through Eqs.
(\ref{ee2}) and (\ref{seq1}), of getting relevant hints which could
lead to the solution
of problem (a) ---at least in a crude way. In the sense that if, once
analyzed in detail, the constraints  (\ref{2sol2}) turned out to be
unphysical, that should lead us to conclude that probably
the operator ordering chosen in the Hamiltonian constraint (and thus the
Wheeler-De Witt equation itself which emerges from it) ought to be
modified.

To be noticed is also the more technical point that, in general,
when performing the analytic continuation
through $s$ (necessary, e.g., for the calculation of determinants
of differential operators), or when
taking the derivative of the series function $F$ with respect
to $s$ at some particular value of $s$, it may well happen
that we hit a pole since, in general, this continuation is a
meromorphic function of $s$. That occurs in our case for
$s=1$. Such a situation can be dealt with in the
usual way, by means of the principal-part prescription
\cite{7,8}. In
Ref. \cite{8} (see also \cite{1}), we discuss several explicit
examples (appearing in
 physical theories) where the precise manner of doing
it is clearly explained, all the way down to the numerical
results (see also \cite{elif3}).

Apart from the physical application that
we have here considered ---directly dealing with
the generalization of the Chowla-Selberg formula derived
above---
in general our new expression will be certainly useful in
physical situations
involving massive theories, finite temperatures or a chemical
potential in a compactified
spacetime. This is the physical meaning to be attributed to the
constant $q$. On
the contrary, for the mathematical uses in number theory the
consideration of
inhomogeneous quadratic forms for defining zeta functions does
not seem
to be specially relevant. This would explain why such formulas
are not to
be found in the mathematical literature, its derivation being,
however, anything but straightforward.

\vspace{5mm}

\noindent{\large \bf Acknowledgments}

I am grateful to Prof. Iver
Brevik for discussions on the Chowla-Selberg
formula, and to
Prof. K{\aa}re Olaussen and to Prof. Lars Brink for the
wonderful hospitality extended to me at
the Universities of Trondheim and G\"{o}teborg, respectively,
where this work was initiated. Enlightening conversations with
Prof. Steven Carlip are gratefully acknowledged. This work
has been partially financed by DGICYT
(Spain), project No. PB90-0022, and by CIRIT (Generalitat de
Catalunya).

\end{document}